\documentclass[journal]{IEEEtran}
\usepackage{amsmath,amssymb,amsfonts}
\usepackage{algorithmic}
\usepackage{graphicx}
\usepackage{textcomp}
\usepackage{wrapfig}
\usepackage{xcolor}
\usepackage{stfloats}
\usepackage{float}
\usepackage{caption}
\usepackage{placeins}
\usepackage{ragged2e}
\usepackage[linesnumbered,ruled,vlined]{algorithm2e} 
\usepackage{booktabs}
\usepackage{subcaption}
\usepackage{cite}
\usepackage[hidelinks]{hyperref}
\usepackage{enumerate}
\begin{document}

	\title{\Huge Movable Antenna-Enabled Integrated Sensing and Communication \\ in Low-Altitude UAV Networks}
	\author{Bin Li, Pengcheng Rao, Xuedong Zhang, and Xinyi Wang,~\IEEEmembership{Member,~IEEE}
		\thanks{(\textit{Corresponding author: Xuedong Zhang.})}
		\thanks{Bin Li and Pengcheng Rao are with the School of Computer Science, Nanjing University of Information Science and Technology, Nanjing 210044, China (e-mail: bin.li@nuist.edu.cn; 202412490770@nuist.edu.cn.)}
		\thanks{Xuedong Zhang is with the School of Computer and Information Engineering, Anhui University of Finance \& Economics, Bengbu 233030, China (e-mail: zxd\_01@163.com).}
		\thanks{Xinyi Wang is with the School of Information and Electronics, Beijing Institute of Technology, Beijing 100081, China  (e-mail: bit\_wangxy@163.com).}
	}
\setlength{\parskip}{0pt}
\maketitle
\begin{abstract}
	This paper investigates a multiple unmanned aerial vehicle (UAV)-assisted integrated sensing and communication (ISAC) system equipped with movable antenna (MA) arrays. To align with practical scenarios, we simulate the dynamic roaming of ground users and the three-dimensional deployment of UAVs in the airspace. We aim to maximize the total data rate by jointly optimizing key operational variables, including UAV trajectories, user association, antenna positions, and beamforming. This formulated problem is subject to constraints on transmission power and the sensing signal-to-noise ratio. To address the challenge of dynamically unknown state transitions due to user mobility, the original problem is decomposed into two steps and solved using different algorithms. First, we utilize the hierarchical density-based spatial clustering of applications with noise (HDBSCAN) algorithm to address the ground-to-air association problem, periodically updating clusters and re-associating during training. The clustering hotspots are used to suggest flight directions for the UAVs. Second, we develop the soft actor-critic algorithm to solve the joint optimization problem of UAV trajectories, antenna positions, and beamforming. Experimental results demonstrate that UAVs equipped with MA arrays outperform those with traditional fixed antenna arrays in ISAC systems, and the proposed optimization strategy effectively enhances communication rates while ensuring sensing performance.
\end{abstract}

	\section{Introduction}
	\label{sec:introduction}
	As the requirements for intelligent terminal devices increase with the advent of sixth-generation mobile communication (6G), the integration of sensing and communication has emerged as an important trend. The integrated sensing and communication (ISAC) systems \cite{1} enable communication and sensing to share spectrum, hardware, and signal processing resources, thus improving the efficiency of spectrum, power, and hardware utilization \cite{2}. ISAC also supports the development of new applications, such as intelligent transportation, autonomous driving, and Internet of Things (IoT) networks. As the demand for high data rates, ultra-reliable connectivity, and precise environmental awareness continues to grow, ISAC has become a key enabling technology for Beyond 5G and 6G wireless systems \cite{111}.\par
	
	Meanwhile, unmanned aerial vehicles (UAV) serving as airborne platforms in wireless networks have attracted extensive attention due to their flexibility and energy efficiency. Owing to their high mobility, adaptive altitude adjustment, and wide coverage capabilities, UAVs can provide more flexible deployment options for dual-functional ISAC base stations compared with terrestrial communication channels \cite{106,108}. In addition, UAVs are also advantageous for sensing tasks, as their aerial deployment provides excellent line-of-sight (LoS) channel conditions, enabling higher sensing accuracy and stability. Moreover, the reflected sensing signals are less likely to be obstructed by obstacles such as buildings \cite{5,107}. \par
	
	Although UAV-assisted ISAC systems have been extensively explored in terms of deployment and resource allocation in recent years, several technical challenges remain  \cite{6}, \cite{7}. First, the joint optimization of UAV trajectory, beamforming, and resource allocation in time-varying wireless environments results in a higher-dimensional and non-convex action decision space. Second, the continuous mobility of UAVs and ground users complicates the synchronization and coordination of beamforming between sensing and communication tasks. Furthermore, traditional fixed-position antennas (FA) are not adaptable to the dynamic migration of UAVs, leading to beam misalignment errors, and the spatial constraints of fixed antennas limit the potential performance improvement of the system \cite{8}. To address these issues, movable antenna (MA) arrays have been proposed. This antenna design overcomes the limitations of traditional fixed-position antennas, enabling the full utilization of spatial degrees of freedom within continuous spatial regions, thereby significantly enhancing the system's flexibility and performance \cite{9}. Compared to FA arrays, MA arrays can dynamically adjust the antenna deployment within the designated transmitter or receiver areas, enabling real-time and optimal utilization of the spatial variations in the wireless channel. Changes in antenna positions also affect the range of the signal radiation angle. At the transmitter, the overall width of the MA array can be adjusted to optimize signal coverage for specific requirements. Additionally, optimizing the geometric structure of the antenna array can reduce the correlation of signals transmitted in different directions, thereby minimizing interference and improving the signal link quality. This adaptability offers several key advantages, including a significant increase in spatial diversity, beamforming gain, and spatial multiplexing performance potential. This allows the MA system to achieve better system performance using the same number of antennas or even fewer \cite{10,11,12}.

The application of MA arrays in wireless communications has been extensively and thoroughly researched. By using different drive components to mechanically alter the antenna positions and orientations, MA designs can be adapted to various specific scenarios. For example, in airborne signal relay frameworks, MA can adaptively deploy antenna positions while enhancing the spatial response of signal reception and transmission directions, thus increasing the system's maximum throughput \cite{1018}. As more potential is explored, wireless sensing and ISAC systems using MA have begun to be widely studied. For time-varying dynamic wireless environments, optimizing the position of MA can compress sensing to improve the accuracy of target parameter estimation, allowing airborne base stations to adapt to changes in channel conditions in dynamic environments \cite{917}.
	
	\subsection{Related Work}
	
	Significant progress has been made in UAV-assisted ISAC research. In \cite{13}, a novel adaptive UAV-assisted ISAC mechanism was proposed, which maximizes the average system throughput by alternately optimizing beamforming and trajectory. The approach also flexibly configures the sensing duration during the UAV's task execution, thereby improving the efficiency of the integrated sensing and communication tasks. The authors of \cite{14} considered a multi-UAV-assisted ISAC system, where they jointly incorporated UAV trajectory and power allocation into the system optimization. They proposed a deep reinforcement learning (DRL) algorithm to maximize detection capabilities while balancing the radar sensing and communication performance of all UAVs. The work in \cite{1022} investigated the real-time trajectory design of UAVs in ISAC systems from a security perspective and formulated a mathematical model for a real-time trajectory design problem. An efficient iterative algorithm was then proposed to obtain an approximate optimal solution. Additionally, in \cite{15}, a joint optimization problem of UAV trajectory and resource allocation in a multi-UAV-assisted ISAC system was proposed. To ensure communication performance while meeting sensing requirements, an alternating optimization mechanism was introduced to solve the non-convex problem and minimize the Cram\'{e}r-Rao lower bound for target detection. 
	
	Based on \cite{13}, \cite{14}, and \cite{15}, it can be observed that in UAV-assisted ISAC systems, UAV trajectory and resource allocation are equally important, as both significantly impact communication and sensing performance. Therefore, the optimization of UAV trajectory and resource allocation has become a key area of current research. Furthermore, in UAV-assisted ISAC systems, the user association problem also affects the performance of both communication and sensing. Xi \textit{et al.} \cite{16} investigated the user association problem in UAV-assisted communication systems and proposed a novel iterative optimization scheme to solve the corresponding mixed-integer non-convex problem, with the objective of maximizing the total communication rate for all users. In \cite{17}, Han \textit{et al.} optimal transport theory was employed to optimize the user association problem in UAV-assisted edge computing networks. Xu \textit{et al.} \cite{1025} proposed an ISAC framework based on a simultaneous transmitting and reflecting reconfigurable intelligent surface with a non-orthogonal multiple access scheme. In their work, user scheduling, active beamforming, and passive beamforming are jointly optimized. Furthermore, a two-layer penalty algorithm was developed to guarantee the user scheduling constraints during the optimization process. Additionally, Dai \textit{et al.} \cite{18} introduced a multi-agent reinforcement learning–based method for jointly optimizing user association and resource allocation. Furthermore, in \cite{5}, Gao \textit{et al.} proposed a K-means-based hierarchical user association algorithm, which dynamically adjusts the binding relationship between UAVs and users during task execution. They then proposed a multi-agent reinforcement learning-based HR-MAPPO algorithm, which jointly optimizes UAV trajectory and beamforming to maximize the total communication rate. 
	
	Despite significant breakthroughs in the research on UAV-assisted ISAC systems, existing studies mainly focus on FA arrays, and the potential of utilizing MA has not been fully explored. In \cite{19}, the authors designed a dual-base radar system integrated into a multi-user MISO ISAC scenario. They achieved array response reconfiguration by adjusting the antenna positions and proposed an optimization problem combining beamforming and antenna positioning. In a multi-user MIMO ISAC scenario, Wang \textit{et al}. \cite{20} designed a fluid antenna system at base station and achieved a higher overall sum rate by optimizing the antenna port positions and beam precoding. Additionally, Chen \textit{et al.} \cite{1033} simulated an ISAC system based on a large MIMO base station supported by MA arrays. By combining three optimization algorithms, they verified that MA can enhance the performance metric range at the transmitter. They also provided a general solution for an ISAC system with MA assistance at both the transmitter and receiver. Most of the research, including the aforementioned work, has effectively validated the performance improvements brought by MA in ISAC systems. However, most of these studies focus on MA designs based on ground base stations, and research on more dynamic scenarios, such as UAVs equipped with MA arrays, is still incomplete. In \cite{21}, Kuang \textit{et al.} investigated an ISAC system with a UAV carrying an MA as an airborne platform, relying on optimizing beamforming and antenna positioning to enhance the beam gain and overall communication rate of the ISAC system. Yu \textit{et al.} \cite{29} investigated the link distortion problem between the ground base station's MA array and the airborne UAV in ISAC systems, and proposed an MA-UAV collaborative control framework that integrates long short-term memory networks to predict the MA positions. 
	
	\subsection{Motivations and Contributions}
	
	The aforementioned studies focus on single-UAV scenarios or neglect the joint optimization of user association, trajectory design, and resource allocation \cite{13}, \cite{14}, \cite{21}. Although existing ISAC research can leverage mature user association techniques from wireless communication systems, the user-UAV association mechanism should be designed for simultaneously
	addressing communication and sensing tasks to enhance performance. Based on the actual sensing requirements and communication tasks, UAVs should strategically select their service targets, which helps avoid the issue of performance degradation caused by excessive utilization of UAV resources \cite{22,1631}. Currently, there is still a lack of research on the joint optimization of UAV trajectories, user association, resource allocation, and antenna positioning with the integration of MA array. 
	
	In UAV-assisted ISAC research, traditional methods such as convex optimization perform well in environments where the number of UAVs and users is limited and static. However, in practical ISAC scenarios, where user positions are typically dynamic and the system becomes more complex as the number of service targets increases, these traditional methods struggle to provide solutions. Given the issue of unknown state transitions caused by users' dynamic movements, this can be formulated as a Markov decision process (MDP). Model-free DRL is well-suited for finding effective policies for MDP in unknown environments through extensive trial-and-error interactions and reward-driven decision iterations \cite{23,24}. To date, DRL has demonstrated powerful capabilities in solving various sequential decision-making problems, such as robot control, gaming, wireless communication, autonomous driving, and large language models. It also exhibits excellent decision-making abilities in UAV-supported ISAC systems. 
	
	Inspired by the aforementioned works, we considered the dynamic deployment of UAVs equipped with MA arrays in 3D space and integrated the MA array into a low-altitude ISAC scenario. Moreover, we use the soft actor-critic (SAC) algorithm in combination with the hierarchical density-based spatial clustering of applications with noise (HDBSCAN) technique to optimize the UAV trajectory, user association, beamforming, antenna positioning, and power allocation. The contributions of this paper are summarized as follows:
	\begin{enumerate}[1)]
		\item We integrate the MA array with the ISAC system and propose a novel low-altitude ISAC system architecture. Unlike traditional fixed-antenna ISAC deployments, the proposed architecture fully exploits the spatial degrees of freedom offered by the MA arrays, enabling dynamic adaptation to the dual task requirements of communication and sensing. Furthermore, to maximize communication performance under sensing constraints, we jointly optimize the MA array deployment, beamforming, user association, and UAV trajectory, subject to the sensing signal-to-interference-plus-noise ratio (SINR) requirements, transmission power budgets, MA mechanical limitations and UAV flight restrictions.
		\item To address the resultant high-dimensional coupled optimization problem, we propose a two-stage solution combining the HDBSCAN clustering algorithm and the SAC algorithm. First, HDBSCAN is used for dynamic clustering of ground users to optimize the air-ground association. This ensures that users associated with the same UAV share similar spatial characteristics, which facilitates subsequent dynamic adjustment of the MA array. Then, based on the clustering results, SAC is applied to solve the joint optimization problem of UAV trajectories, antenna deployment, and beamforming.
		\item We experimentally validate the proposed scenario and algorithm framework. The results show that the integration of the MA array with the ISAC system significantly improves the total communication rate compared to traditional fixed-antenna systems, while more stably meeting the sensing SINR constraints. Furthermore, through a step-by-step comparison of optimization components, we quantify and validate the contribution of the MA array to the performance of the ISAC system.
	\end{enumerate}
	
	The remainder of this paper is structured as follows. Section \ref{section:2} elaborates on the system model and the formulation of the optimization problem. Section \ref{section:3} details the user association strategy and the procedural specifics of the proposed algorithm. Section \ref{section:4} presents and discusses the experimental results. Finally, Section \ref{section:5} concludes this paper.
	
	\section{System Model}
	\label{section:2}
\begin{figure}[t]
		\centering
		\includegraphics[width=0.5\textwidth]{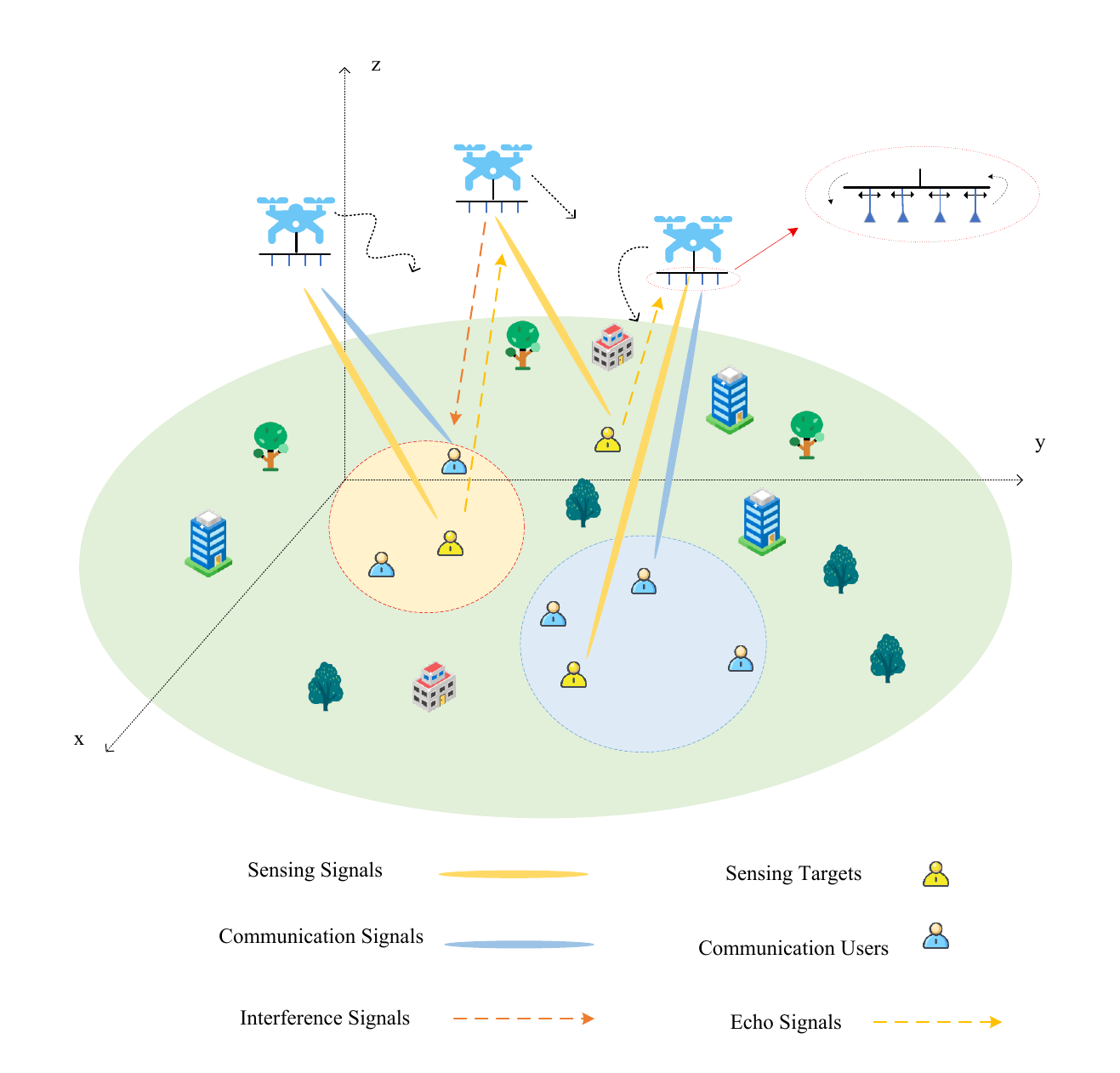}
		\caption{Illustration of the UAV-enabled ISAC system.}
		\label{t1}
	\end{figure}

	As shown in Fig. \ref{t1}, we consider a multi-UAV-enabled ISAC system where UAVs are deployed in the near-airspace to act as dual-functional base stations. They provide downlink communication services to ground users while simultaneously performing radar sensing tasks for sensing targets. We assume that each UAV is equipped with an MA array, and each user is equipped with a single receive antenna. In this scenario, a 3D Cartesian coordinate system and a finite mission duration are considered. At each time slot $t\in {\large \mathcal{T} }  = \left\{ 1,...,T \right \} $, the location of UAV $n\in \mathcal{N}  =\left \{ 1,...,N \right \}  $ is denoted as $\textbf{L}_{n}[t]=(x_{n}[t],y_{n}[t],z_{n}[t]) $, while communication user  $k_{c} \in \mathcal{K}_{c}=\left\{1,...,K_{c}\right\}$  and sensing target $k_{s}\in\mathcal{K}_{s}=\left\{1,...,K_{s}\right\}$  are positioned at $\textbf{L}_{k_{c}}[t]=(x_{k_{c}}[t],y_{k_{c}}[t],0)$ and $\textbf{L}_{k_{s}}[t]=(x_{k_{s}}[t],y_{k_{s}}[t],0)$. Meanwhile, to represent the association between UAVs and ground users, a binary variable $\alpha_{n,k}[t]$ is introduced. If user $k$ is associated with UAV $n$ in time slot $t$, then $\alpha_{n,k}[t]=1$, otherwise, $\alpha_{n,k}[t]=0$. Within the same time slot, each user can be associated with only one UAV, while each UAV can serve multiple users simultaneously.  \par
	Moreover, user mobility is incorporated into this scenario. The speed and movement direction of each user are randomly assigned within a reasonable range. To further enhance the randomness of movement, both the speed and direction are periodically resampled at predefined time intervals.  
	
	\subsection{UAV Trajectory Model}
	
	In most existing studies where UAVs serve as aerial ISAC platforms, they are typically deployed at a constant altitude. To better align with practical deployment scenarios, this paper investigates the motion of UAVs in 3D space. For any UAV, its position at any time is constrained within a fixed range, i.e.,
		$x_{\mathrm{min}}\le x_{n}[t]\le x_{\mathrm{max}},$ 
		$y_{\mathrm{min}}\le y_{n}[t]\le y_{\mathrm{max}},$ and
		$z_{\mathrm{min}}\le z_{n}[t]\le z_{\mathrm{max}},$
	where $(x_{\mathrm{min}},y_{\mathrm{min}},z_{\mathrm{min}})$ and $(x_{\mathrm{max}},y_{\mathrm{max}},z_{\mathrm{max}})$ represent the minimum and maximum values of the UAV's operational range, respectively. Moreover, the maximum speed of the UAV is constrained not to exceed $\mathit{V}$ at any time, expressed as 
	\begin{equation}
		\left \| \textbf{L}_{n}[t+1]-\textbf{L}_{n}[t] \right \|\le V ,~\forall n\in \mathcal{N}.
	\end{equation}
	Due to the maximum speed limitation, the UAV’s movement distance within each time slot is limited. Let $\nabla\textbf{L}_n[t]=v_n[t]\textbf{q}_n[t]$ represent the position change of UAV $n$ during time slot $t$, where $v_n[t]$ and $\textbf{q}_n[t]$ denote the flight speed and the unit vector of the flight direction generated by the agent, respectively. Therefore, the UAV trajectory model can be expressed as
	\begin{equation}
    	\textbf{L}_n[t]=\textbf{L}_n[0]+\sum\limits_{t^{'}=1}^{t}\nabla\textbf{L}_n[t^{'}],~\forall n\in \mathcal{N}.
	\end{equation}
	
	Additionally, to prevent trajectory conflicts and potential collisions between UAVs, a minimum safe distance $D_{0}$ must be maintained between any two UAVs at all times. This constraint is expressed as 
	\begin{equation}
		\left \| \textbf{L}_{m}[t]-\textbf{L}_{n}[t] \right \|\ge D_{0} ,~\forall n,m\in \mathcal{N},n\not= m.
	\end{equation}
	
	\subsection{Antenna Model}
	Each UAV is equipped with an MA array consisting of $M$ antennas, where the set of MA is given by $\mathcal{M}=\left\{1,...,M\right\}$. We assume that the MA array can adjust its position along a fixed-length one-dimensional segment, with each antenna capable of continuous position changes according to practical requirements. The relative offset position of each antenna element in the array of UAV $n$ with respect to the array center at time slot $t$ can be denoted by $ \textbf{r}_{n}(t)=[\textbf{r}^n_{1}(t),\textbf{r}^n_{2}(t),...,\textbf{r}^n_{M}(t)]^{T}$, where $-D_{\textrm{off}}\leq\textbf{r}^n_{1}(t)<\textbf{r}^n_{2}(t)<...<\textbf{r}^n_{M}(t)\leq D_{\textrm{off}}$, and any two antennas keep a minimum distance $D_{\textrm{min}}$. The MA system discussed in \cite{10,11,12} typically employs miniature motors or piezoelectric actuators, which can precisely control the antenna positions within a millimeter-scale range. These antennas, mounted on mechanical structures within the wavelength-level sliding rail range, are well within the UAV's payload capacity. Therefore, the above antenna design remains feasible within the practical mechanical constraints. Furthermore, unlike the MA arrays deployed at terrestrial macro base stations, the movable range of MA elements mounted on UAVs is more limited. As such, the velocity of existing micro linear actuators suffices for task execution within second-level time slots. Additionally, the energy loss caused by antenna movement is much smaller than the UAV's flight energy consumption, and it is considered secondary in terms of improving communication and sensing performance. Thus, the energy consumption of the MA is not explicitly included in the optimization objective of this scenario.
	
	Since UAV flight inevitably involves vibrations and attitude variations that lead to deviations in the array’s orientation, we assume that the segment on which the array lies is allowed to rotate freely. When the UAV is hovering normally, the direction of the antenna array is $\vec{\textbf{v}}=(\cos\iota,\sin\iota,0)$, where $\iota$ represents the angle between the array and the x-axis. We introduce $\{\phi ,\psi, \varsigma \}$ to model the UAV’s attitude deviation. Specifically, $\phi$, $\psi$ and $\varsigma$ denote the roll angle, pitch angle, and yaw angle, respectively, corresponding to the UAV's rotational angles about the $x$-axes, $y$-axes and $z$-axes. The array rotation caused by the UAV’s attitude deviation can be defined as
	\begin{align}
		&\textbf{R}^n = \textbf{R}^n_z(\varsigma) \textbf{R}^n_y(\psi) \textbf{R}^n_x(\phi) =\notag\\
		& \scalebox{0.8}{$
			\begin{bmatrix}
				\cos \psi \cos \varsigma & \cos \varsigma \sin \psi \sin \phi - \sin \varsigma \cos \phi & \cos \varsigma \sin \psi \cos \phi + \sin \varsigma \sin \phi\\
				\cos\psi \sin \varsigma  & \sin \varsigma \sin \psi\sin \phi  + \cos \varsigma \cos \phi & \sin \varsigma \sin \psi\cos \phi - \cos \varsigma \sin \phi\\
				-\sin \psi & \sin \phi \cos \psi & \cos \phi\cos\psi
			\end{bmatrix}$
		},
	\end{align}
	where $\textbf{R}^n_z(\varsigma), \textbf{R}^n_y(\psi)$ and $\textbf{R}^n_x(\phi)$ represent the rotation matrices for the three directional components.
	
	Based on the above definitions, in order to ensure that the MA array always remains aligned with a plane parallel to the ground, the angles of active rotation required to counteract UAV attitude deviations can be derived as  
	\begin{equation}
		\mu=f(P,Q)=\begin{cases}
			\arctan \frac{P}{Q},  & Q>0 \\
			\arctan \frac{P}{Q}+\pi,  & P\ge 0,Q<0 \\
			\arctan \frac{P}{Q}-\pi, & P<0,Q<0\\
			~~\frac{\pi}{2},  & P>0,Q=0\\
			- \frac{\pi}{2}, & P<0,Q=0
		\end{cases}~~,
	\end{equation}
	where $P=-\textbf{R}^n\vec{\textbf{v}}\textbf{e}_z$ and $Q=(\textbf{R}^n \vec{\textbf{l}})\times(\textbf{R}^n\vec{\textbf{v}})\textbf{e}_z$, with  $\textbf{e}_z$ being the unit vector in the $z$-axis direction, and $\vec{\textbf{l}}$ represents the direction of the array's rotation axis, which aligns with the $z$-axis when the array is parallel to the ground\footnote{The rotation compensation mechanism is built based on rigid-body mechanics and vector projection theory, which provides a clear physical and mathematical foundation. The attitude deviation angles required in the rotation matrix derivation can be obtained in real time from the inertial measurement unit at low cost.}. Attitude deviations caused by airflow disturbances and UAV maneuvering are generally minor; consequently, the antenna array does not require large rotations, and the required adjustment range is fully within the capabilities of current industrial control units. Although some delay may occur between beam transmission and array rotation in more practical scenarios, this study focuses on the theoretical performance gain of the MA array under ideal conditions, and thus practical error factors are not considered. Therefore, we can determine that the array is spread out within the ideal plane. Let $\theta_{n,k_c}[t]$ denote the physical steering angle of user $k_c$ with respect to UAV $n$ when the array is parallel to the ground. Since the distance between the UAV and user is typically much larger than the size of the region for antenna movement, we consider that the steering vector of the MA array can be expressed as
	\begin{align}
		\textbf{a}_{n,k_c}[t]=[e^{j\frac{2\pi }{\lambda }r^n_{1}[t]\cos \theta _{n,k_c}[t]}, ... , e^{j\frac{2\pi }{\lambda } r^n_{m}[t]\cos \theta _{n,k_c}[t]}]^T,
	\end{align}
	
	\noindent where $\lambda$ is the wavelength. 
	\subsection{Communication Model}
	
	Due to the high altitude of the UAVs, we assume that the air-to-ground link between UAV $n$ and user $k_{c}$ in time slot $t$ is dominated by a line-of-sight (LoS) channel, and the channel power gain follows the free-space path loss model. Therefore, the channel gain can be expressed as 
	\begin{equation}
		\textbf{h}_{n,k_{c}}[t]=\sqrt{\beta_{0}d^{-2}_{n,k_{c}}[t]}\textbf{a}_{n,k_c}[t],\label{eq:3}
	\end{equation}
	where $\beta_{0}$ is the reference channel gain when the distance is 1 m,  $d_{n,k_{c}}[t]=\sqrt{(x_{n}[t]-x_{k_{c}}[t])^2+(y_{n}[t]-y_{k_{c}}[t])^2+z^2_{n}[t])}$ denotes the distance between UAV $n$ and user $k_{c}$ at time slot $t$.
	
	In this scenario, the UAV may be associated with multiple users. Due to the limited resources of the UAV, we adopt discrete control over the transmit power allocated to each associated user. Assuming the maximum transmit power of the UAV is $\mathrm{P}_{\mathrm{max}}$, the power allocation ratios for the UAV to each associated user at time slot $t$ are represented by a vector $\boldsymbol{\rho}_{n}[t]=[\rho^n_{1}[t],...,\rho^n_{k_{c}}[t]]$, where $\rho^n_{k_{c}}[t]\le1$. Note that the transmit power allocation ratios for the UAV to non-associated users are zero. Then, the transmit power from the UAV to each user at each time slot $t$ can be expressed as
	\begin{equation}
		\textbf{P}_{n,k}[t]=|\rho^n_{k}[t]\textbf{w}_{n,k}[t]|^2,\forall k\in \mathcal{K}_{c},
	\end{equation}
	where $\textbf{w}_{n,k}[t] \in \mathbb{C}^{M\times 1}$ denotes beamforming vector from UAV $n$ to user $k$ at time slot $t$.
	
	Based on the above definitions, the signal transmitted by UAV $n$ at time slot $t$ can be expressed as 
	\begin{equation}
		\textbf{x}_{n}[t]=\sum_{k}^{K_{c}}\alpha_{n,k}[t]\rho^n_{k}[t]\textbf{w}_{n,k}[t]\textbf{s}_{n,k}[t],
	\end{equation}
	where $\textbf{s}_{n,k}[t]\in \mathbb{C}^{1\times L}$ denotes the symbol vector from UAV $n$ to user $k$ at time slot $t$. Meanwhile, the received signal at user $k$ at the same time slot can be expressed as follows: 
	\begin{align}
		\textbf{y}_{k}[t]=\sum\limits_{n=1}^{N}&\alpha_{n,k}[t]{\bigg (}\textbf{h}^H_{n,k}[t]\rho^n_{k}[t]\textbf{w}_{n,k}[t]\textbf{s}_{n,k}[t]\notag\\
		&+\sum\limits_{l\not= k}\textbf{h}^H_{n,l}[t]\rho^n_{l}[t]\textbf{w}_{n,l}[t]\textbf{s}_{n,l}[t]\notag\\
		&+\sum\limits_{m\not= n,l\not= k}^{}\textbf{h}^H_{m,l}[t]\rho^m_{l}[t]\textbf{w}_{m,l}[t]\textbf{s}_{m,l}[t]+n_0\bigg), \label{eq:7}
	\end{align}
	
	\noindent where $n_{0}$ represents the additive white Gaussian noise (AWGN). \par
	As shown in (\ref{eq:7}), the received signal at the user $k$ includes not only the communication signal but also interference caused by transmissions from the associated UAV $n$ to other users, as well as interference from non-associated UAVs. The power of these interference signals can be expressed as
	\begin{align}
		\textbf{I}_k=\sum\limits_{l\not=k}\ \big|\textbf{h}^H_{n,l}[t]\rho^n_{l}[t]&\textbf{w}_{n,l}[t]\big|^2\notag\\
		&+\sum\limits_{l\not=k,m\not=n}\ \big|\textbf{h}^H_{m,l}[t]\rho^m_{l}[t]\textbf{w}_{m,l}[t]\big|^2.
	\end{align}
	Thus, the SINR for user $k$ at time slot $t$ is defined as
	\begin{equation}
		\gamma _{k}[t]=\sum\limits_{n=1}^{N}\alpha _{n,k}[t] \frac{\ \big| \textbf{h}^H_{n,k}[t]\rho^n_{k}[t]\textbf{w}_{n,k}[t]\big |^2}{\textbf{I}_k+\sigma ^2_{c}},
	\end{equation}
	\noindent where $\sigma^2_{c}$ represents the power of the AWGN. Therefore, the maximum achievable communication rate for user $k$ can be expressed as
	\begin{equation}
		R_{k}[t]=B\log_{2}{(1+\gamma _{k}[t])}, 
	\end{equation}
	
	\noindent where $B$ represents the bandwidth.
	
	\subsection{Sensing Model}
	
	In the sensing process, the UAV needs to transmit sensing signals to the target and receive the corresponding echo signals to achieve the sensing function. The core of sensing performance lies in the reception and processing of echo signals. Compared to ground systems, UAV systems are more sensitive to interference and noise in the echo signals. In most previous ISAC systems supported by multiple UAVs, the sensing task is typically performed by a single UAV, which significantly limits the improvement of sensing performance. In this system, we adopt a distributed sensing approach for a single target using multiple UAVs: one UAV is designated as the transmitter to emit sensing signals, while another UAV serves as the receiver to capture the echo signals. To ensure consistency of the association within a time slot, the transmitter remains the UAV previously associated with the target. The selection of the receiver, however, is based on the cluster point assigned to each UAV. If two UAVs are responsible for the same cluster, the target’s sensing signal is received by another non-associated UAV. If each cluster point is handled by a single UAV, the receiver is chosen as the non-associated UAV that is closest and responsible for the cluster with the lowest user density.
	
	We set the sensing and communication tasks to be performed in separate time slots to reduce the interference of communication signals on sensing. The channel response matrix between the transmitter UAV $n$ and the receiver UAV $j$, relative to the sensing target $k_s$, can be expressed as 
	\begin{equation}
		\textbf{H}^s_{n,j,k_s}[t]=\varrho^{k_s}_{n,j}[t]\textbf{a}_{n,k_s}[t]\textbf{a}^H_{j,k_s}[t]~,\label{eq:11}
	\end{equation}
	where $\varrho_{n,j}^{k_s}=\sqrt[]{\frac{\kappa^2\epsilon _{k_s}}{d^2_{n,k_s}[t]d^2_{j,k_s}[t]}}$, with $\epsilon_{k_s}$ denotes the radar cross-section of target $k_s$ and $\kappa$ is the channel power at the reference distance. During task execution, the transmitted signal may encounter clutter objects on the ground, leading to signal scattering. As the receiver captures the echo signal, it typically also receives clutter signals generated by these scatterers. The interference from clutter signals reduces the sensing SINR, which in turn affects the target detection probability. In user-dense scenarios, clutter can become a major factor limiting sensing performance. In this context, other non-target users located within the ground elliptical region defined by the transmitter and receiver as the focal points are regarded as scattering points, and the signals reflected by these users are considered as clutter interference. The impact of clutter interference also indirectly reflects the effect of high communication user density on sensing performance. Let $\textbf{s}^s_n[t]$ represent the sensing signal transmitted by the transmitter UAV $n$. Then, the echo signal received by the receiver UAV $j$ from the  transmitter UAV $n$ at time slot $t$ can be expressed as
	\begin{equation}
		x^r_{j}[t]= \sum_{n=1}^{N}\alpha_{n,k_s}[t]\big(\textbf{H}^s_{n,j,k_s}[t]\textbf{w}_{n,k_s}[t]\textbf{s}^s_{n}[t-\tau_{n,j}]+\textbf{I}_j[t]+\textbf{n}_j[t]\big),\label{eq:rec}
	\end{equation}
	where $\tau_{n,j}$ denotes the transmission delay, and $\textbf{I}_j[t]=\sum\limits_{k_c}^{\mathcal{C}_k}\varrho_0\textbf{a}_{n,k_c}\textbf{a}^H_{j,k_c}\textbf{w}_{n,k_s}[t]\textbf{s}^s_{n}[t-\tau_{n,j}]$ represents the clutter signal caused by the scattering of the sensing signal from other users within the sensing region, with $\varrho_0$ is random scattering coefficient of the user. Based on the signal model in (\ref{eq:rec}), we can apply matched filtering to the echo signal, thereby distinguishing the useful signal from the clutter in the mixed signal. Meanwhile, we define $\textbf{u}_{j,k_s}[t]=\frac{\textbf{a}_{j,k_s}[t]}{\left \| \textbf{a}_{j,k_s}[t] \right \| _2}$ as the receive beamforming vector, which is applied at the receiver to enhance signal reception from the target direction while suppressing clutter interference and environmental noise.\par
	Based on the above definition, the sensing SINR of the sensing target $k_s$ can be defined as  
	\begin{equation}
		\Gamma _{k_{s}}[t]=\sum_{n=1}^{N}\alpha_{n,k_s}[t]\frac{\big|\textbf{u}^H_{j,k_s}[t]\textbf{H}^s_{n,j,k_s}[t]\textbf{w}_{n,k_s}[t]\big|^2}{\sum\limits_{k_c}^{\mathcal{C}_k}\big|\textbf{u}^H_{j,k_s}[t]\varrho_0\textbf{a}_{n,k_c}\textbf{a}^H_{j,k_c}\textbf{w}_{n,k_s}[t]\big|^2+\sigma^2_{s}},
	\end{equation}
	where $\sigma^2_{s}$ is the power of the AWGN. Given a fixed false alarm rate, there exists a positive correlation between the radar detection probability and the sensing SINR. In the considered ISAC system, the sensing SINR is used as the indicator of sensing performance\footnote{In this scenario, our objective is to enhance communication performance while guaranteeing reliable sensing. The core requirement of the sensing task lies in the quality of sensing signal transmission and reception. Moreover, this work focuses on exploring the potential of MA arrays in a multi-UAV ISAC system, and thus the sensing performance metric should directly reflect the impact of antenna position optimization. Accordingly, in the considered ISAC scenario, the SINR defined above is adopted as the sensing performance metric.}. During task execution, we require that the SINR always remains above the predefined minimum threshold to ensure sufficient sensing strength throughout the task.
	
	\subsection{Problem Formulation}
	
	We formulate the problem as maximizing the overall communication sum rate by jointly optimizing the UAV trajectory, user association, beamforming design, and antenna array deployment, subject to the UAV flight constraints, sensing quality requirements, and the maximum transmit power limitation, while also considering the UAV resource allocation. Therefore, the optimization problem can be expressed as
	\setcounter{equation}{16} 
	\begin{subequations}\label{eq:18}
		\begin{align}
			\mathcal{P}1: \quad
			& \max_{\textbf{L}_{n},\alpha_{n,k},\textbf{w}_{n,k},\textbf{r}_n,\rho^n_{k}} \quad\sum_{k=1}^{K_c} \sum_{t=1}^{T} R_k[t], \notag \\
			\text{s.t.} \quad
			&\textbf{P}_{n,k}[t] \leq \mathrm{P}_{\mathrm{max}}, ~ \forall k \in \mathcal{K}_c,~ \forall n \in \mathcal{N}, \label{eq:18a} \\
			&\Gamma_{k_s}[t] \geq \Gamma_{\mathrm{thr}}, ~\forall k \in \mathcal{K}_s, \label{eq:18b} \\
			&|\textbf{r}^n_{m}| \leq D_{\mathrm{off}}, 
			~ \forall n \in \mathcal{N}, ~ \forall m \in \mathcal{M}, \label{eq:18c} \\
			&|\textbf{r}^n_{m+1}-\textbf{r}^n_{m}| \geq D_{\mathrm{min}}, 
			~ \forall n \in \mathcal{N}, ~ \forall m \in \mathcal{M}, \label{eq:18d} \\
			&x_{\mathrm{min}} \leq x_n[t] \leq x_{\mathrm{max}},~ \forall n \in \mathcal{N},\label{eq:18e}\\
			&y_{\mathrm{min}} \leq y_n[t] \leq y_{\mathrm{max}},~ \forall n \in \mathcal{N},\label{eq:18f}\\ 
			&z_{\mathrm{min}} \leq z_n[t] \leq z_{\mathrm{max}}, ~ \forall n \in \mathcal{N}, \label{eq:18g} \\
			&\|\textbf{L}_m[t] - \textbf{L}_n[t]\| \geq D_0, 
			~ \forall n,m \in \mathcal{N}, ~ n \neq m, \label{eq:18h} \\
			&\|\textbf{L}_n[t+1]-\textbf{L}_n[t]\| \leq V, ~ \forall n \in \mathcal{N}, \label{eq:18i} \\
			&\sum_{n=1}^N \alpha_{n,k}[t] = 1, 
			~ \forall k \in (\mathcal{K}_c \cup \mathcal{K}_s). \label{eq:18j}
		\end{align}
		
	\end{subequations}
	\noindent Constraint (\ref{eq:18a}) indicates that the communication signal transmission power of the UAV must not exceed the rated maximum value $\mathrm{P}_{\mathrm{max}}$ at any time. Constraint (\ref{eq:18b}) ensures that the SINR of the sensing signal remains above the threshold $\Gamma_{\mathrm{thr}}$. Constraints (\ref{eq:18c}) and (\ref{eq:18d}) describe the mechanical limitations of antenna movement. Here, $D_{\mathrm{off}}$ represents the maximum offset distance of an antenna from the array center, which also corresponds to half the length of the MA array. Additionally, the minimum distance $D_{\mathrm{min}}$ between any two antennas must be maintained, defined as half the wavelength. This distance is the minimum required to prevent inter-element coupling in a uniform scattering environment. Constraint (\ref{eq:18e}), (\ref{eq:18f}) and (\ref{eq:18g}) define the deployment range limitation of the UAV. Constraint (\ref{eq:18h}) requires that a safe collision avoidance distance $D_0$ be maintained between any two UAVs. Constraint (\ref{eq:18i}) imposes a flight speed limit on the UAV. Constraint (\ref{eq:18j}) stipulates that a user can only be associated with one UAV at any given time. 
	
	\section{Proposed Solution}
	\label{section:3}
	Problem ($\mathcal{P}$1) is a multivariable, complex, and non-convex optimization problem, where the decision variables are highly coupled. To address this challenge, we decompose ($\mathcal{P}$1) into two subproblems: (i) the optimization of user association, and (ii) the joint optimization of communication and sensing beamforming, UAV trajectory, movable antenna array configuration, and power allocation. For the first subproblem, we adopt an HDBSCAN-based clustering algorithm to determine the user association. For the second subproblem, we employ the SAC algorithm to solve the remaining joint optimization tasks. 
	
	\subsection{HDBSCAN-Based User Association}
	
It is note that we need a more stable air-ground association strategy to ensure the array is deployed effectively and efficiently. If we base the air-ground association on distance, the relationship would change frequently as the UAVs move, leading to constant adjustments of the MA array. This would reduce the performance gains from the optimization, as system performance would be more affected by the UAV trajectory, which is not the focus of our study. To address this issue, we employ the HDBSCAN clustering algorithm for association. During training, we periodically cluster ground objects and associate them with UAVs based on the clusters. This approach ensures that the objects served by each UAV have more similar spatial distribution, making the adjustment of the MA array based on these clusters more efficient and reasonable.\par
	\begin{algorithm}[t]
		\caption{HDBSCAN-Based User Association Algorithm}
		\label{alg:HDBSCAN}
		
		\KwIn{UAV number $N$, UAV locations $\textbf{L}_n$, communication users' locations $\textbf{L}_{kc}$, association interval $T_c$, maximum timestep $T$, maximum UAV load $\xi$.}
		\KwOut{UAV--user association and UAV flight trajectories.}
		
		Initialize HDBSCAN cluster with parameters min\_cluster\_size, min\_samples, and $\epsilon$.
		
		\For{\textrm{each timestep} $t < T$}{
			\If{$t \bmod T_c = 0$}{
				Apply HDBSCAN on $\{\textbf{L}_{kc}\}$ to obtain clusters $\{\textbf{C}_1,\dots,\textbf{C}_U\}$ and ignore outliers.\\
				Compute cluster centroids $\boldsymbol{\mu}_u = \frac{1}{|\textbf{C}_u|}\sum_{\textbf{L}_{kc}\in \textbf{C}_u}\textbf{L} _{kc}$.\\
				Construct distance matrix $d_{nu} = \|\textbf{L}_n - \boldsymbol{\mu}_u\|$.\\
				\eIf{$U \geq N$}{
					Use Hungarian algorithm to assign UAVs to clusters by minimizing $\sum d_{nu}$.
				}{
					Assign each UAV $n$ to the nearest cluster $\arg\min_{u} d_{nu}$.
				}
			}
			\For{each UAV $n=1,\dots,N$}{
				 Update the association $\alpha_{n,k}[t]$ based on the UAV's assigned clustering $\boldsymbol{\mu}_{u(n)}$ and users' relative position.
				}
		}
	\end{algorithm}
	The HDBSCAN algorithm is a density-based hierarchical clustering method. Its advantage lies in not requiring a predefined number of clusters and not forcing outlier users to be assigned to any cluster. Sample points within the same cluster have more similar spatial characteristics, which makes HDBSCAN more robust in ISAC scenarios. In this optimization framework, we periodically apply the HDBSCAN algorithm to perform clustering analysis of ground users and follow a specific strategy to assign cluster points to each UAV. Users not included in any cluster will be associated with UAVs using a secondary coverage strategy. The detailed algorithmic procedure is presented in \textbf{Algorithm \ref{alg:HDBSCAN}}.
	
	\subsection{DRL-Based Beamforming, Trajectory, Movable Antenna Array Configuration, and Power Allocation Optimization}
	
	After obtaining the user association from the first-stage clustering algorithm, the agent subsequently makes decisions on beamforming, UAV trajectory, antenna array adjustment, and power allocation according to the current environment. To model this optimization problem, we represent it as an MDP. The MDP is defined by the tuple $\{\mathcal{S},\mathcal{A},\mathcal{P},\mathcal{R}\}$,  where $\mathcal{S}$ is the state space, $\mathcal{A}$ is the action space, $\mathcal{P}$ is the state transition function, and $\mathcal{R}$ is the reward function.
	
	In our scenario, the agent needs to observe the channel information and user locations to design the beamforming and deploy the antenna positions based on the environmental state. Therefore, the state $s_t \in \mathcal{S}$ is given by
	\begin{equation}
		s_t=\{\textbf{h}_{n,k_c}[t],\textbf{H}^s_{n,j,k_s}[t],\textbf{L}_{\textrm{UAV}}[t],\textbf{L}_{\textrm{user}}[t],\textbf{r}^{\textrm{pre}}[t]\},
	\end{equation}
	where $\textbf{L}_{\textrm{UAV}}[t]$ represent the positions of UAVs, $\textbf{L}_{\textrm{user}}[t]$ denotes the position of all users and targets, and $\textbf{r}^{\textrm{pre}}[t]$ refers to the current deployment of each antenna array.
	After observing the environmental state, the agent takes actions based on the state. The actions output by the policy include beamforming design, UAV trajectory, antenna deployment, and power allocation. Therefore, the action $a\in \mathcal{A}$ at time $t$ can be expressed as
	\begin{equation}
		a_t=\{\textbf{w}[t],\boldsymbol{\rho}_{n}[t],\textbf{V}_{\textrm{UAV}}[t],\textbf{C}_{\textrm{UAV}}[t],\textbf{r}[t]\},
	\end{equation}
	where $\textbf{V}_{\textrm{UAV}}[t]$ and $\textbf{C}_{\textrm{UAV}}[t]$ represent the expected flight speed and direction of the UAV. 
	The reward function is designed based on our optimization objectives, where the instantaneous reward $r_t$ at each time step is defined as follows:
	\begin{equation}
		r_t=\sum^{K_c}_{k}R_k[t]-f[t],
	\end{equation}
	where $f[t]$ represents the accumulated penalty when the sensing SINR and other constraints are not satisfied during the UAV’s operation. The penalty for any sensing target not meeting the sensing requirements is denoted as $\frac{\Gamma_{\textrm{thr}}}{\Gamma_{k_s}[t]}f_1$, with a larger penalty applied when the sensing SINR is smaller. When the distance between any two UAVs is smaller than the safe collision avoidance distance $D_0$, a fixed penalty is added. A penalty $f_2$ is also incurred if the UAV speed exceeds the threshold $V$. Both $f_1$ and $f_2$ represent the penalty weights for the corresponding constraint violations. \par
	
	Subsequently, we treat the UAV as the agent interacting with the environment and apply the SAC algorithm to solve the MDP formulated above. As an off-policy DRL algorithm, SAC exhibits notable advantages over conventional approaches in terms of sample efficiency, training stability, and exploration capability. Built upon the actor–critic framework, the actor module aims to maximize both the expected cumulative reward and the policy entropy, while the critic module evaluates the performance of the current policy. By incorporating an entropy regularization term into the objective function, SAC balances reward maximization with entropy maximization. This approach enhances the agent’s exploration capability and effectively prevents convergence to suboptimal local solutions caused by insufficient exploration \cite{32}. The objective function of SAC can be expressed as \cite{33}
	\begin{equation}
		J(\pi) = \mathbb{E}_{\pi} \left[ \sum_{t=0}^{T} \gamma^{t} \big( r_t + \lambda_e H(\pi(\cdot \mid s_t)) \big) \right],
	\end{equation}
	where $\pi$ denotes the policy and policy entropy $H(\pi(\cdot|s ))=-\mathbb{E}_{a_t:\pi(\cdot|s)}[\log_{}{\pi(a_t|s_t)} ] $. $\gamma\in[0,1]$ is the discount factor, which indicates the agent's balance between immediate rewards and future rewards during training. $\lambda_e$ is the entropy coefficient, which controls the weight of the entropy term. A larger value of $\lambda_e$ encourages the agent to take more diverse actions in each state $s_t$, promoting exploration during the training process.\par
	In the SAC algorithm, the critic component consists of two Q-networks, $Q_{\theta_1}(s_t,a_t)$ and $Q_{\theta_2}(s_t,a_t)$, and two target Q-networks, $Q_{\theta^{'}_1}(s_t,a_t)$ and $Q_{\theta^{'}_2}(s_t,a_t)$. Each network computes the Q-value and target Q-value for the action $a_t$ taken at each state, representing the expected return for the action in that state. When updating the two Q-networks at a given frequency, the algorithm selects the minimum value between the two target Q-values. This approach helps to provide a more accurate estimation of the Q-values and mitigates the problem of Q-value overestimation. We denote the learning rate of the Q-networks as $\alpha_c(t)$. The target values for the update of the two Q-networks are given by the following equation \cite{32} 
	\begin{align}
		y_t = r_t + \gamma \mathbb{E}_{s_{t+1}:\mathcal{P}}[ \mathbb{E}_{a_{t+1} \sim \pi_\vartheta } [ &\min\limits_{\small{i=1,2}} {Q_{\theta^{'}_i} (s_{t+1}, a_{t+1})}   \notag\\
		&- \lambda_e \log \pi_\vartheta (a_{t+1} | s_{t+1}) ]]~,\label{eq.Qv}
	\end{align}
	where $\pi_\vartheta$ denotes the policy network. We define the loss function for updating the Q-networks as \cite{33} 
	\begin{equation}
		L(\theta_i) = \mathbb{E}_{(s_t, a_t) \sim \mathcal{M}} \left[ \frac{1}{2} \left( Q_{\theta_i}(s_t, a_t) - y_t \right)^2 \right],\label{eq:Qpra}
	\end{equation}
	where $\mathcal{M}$ denotes the experience replay buffer, which is used to store the agent's past experiences in the form of state-action-reward transitions. These experiences are sampled and used during training to break temporal correlations and improve the stability of the learning process.\par
	The actor component of the algorithm consists of a policy network $\pi_\vartheta(a_t|s_t)$, which is responsible for selecting actions based on the current state. During the training process, the policy network is updated to maximize the expected Q-value and the entropy term. This ensures that the agent's policy retains a balance between exploitation and exploration, preventing premature convergence to suboptimal solutions. We define the loss function for updating the policy network as \cite{32} 
	\begin{equation}
		L(\vartheta) = \mathbb{E}_{s_t \sim \mathcal{M}} [ \mathbb{E}_{a_t \sim \pi_\vartheta} [ \lambda_e \log \pi_\vartheta(a_t | s_t) - \min_{i=1,2} Q_{\theta_i}(s_t, a_t) ]].\label{eq:Ppra}
	\end{equation}
	The loss function for updating the entropy coefficient $\lambda_e$ is defined as \cite{33} 
	\begin{equation}
		L(\lambda_e) = \mathbb{E}_{a_t \sim \pi_\vartheta} \left[ -\lambda_e \log \pi_\vartheta(a_t | s_t) - \lambda_e H_{\text{target}} \right],\label{eq:entc}
	\end{equation}
	where $H_\text{target}$ represents the target entropy, which is typically set to $-dim(a_t)$.\par
	The detailed training process of the SAC algorithm is provided in \textbf{Algorithm \ref{alg:SAC}}. 
	
	\begin{algorithm}[t]
		\caption{SAC-Based Beamforming, Movable Antenna Array Configuration, and Power Allocation Optimization}
		\label{alg:SAC}

		Initialize policy network parameters $\vartheta$, Q-networks parameters $\theta_1$, $\theta_2$, target Q-networks parameters $\theta^{'}_1$, $\theta^{'}_2$, and entropy coefficient $\lambda_e$. \\
		Initialize actor network, critic networks, optimizers for each network. \\
		Initialize experience replay buffer $\mathcal{M}$.

		\For{epoch e$= 1$ to $N_{\text{epoch}}$}
		{
			Obtain the initial state $s_1$ of the environment.
			
			\For{time step t$= 1$ to $T_{\text{step}}$}{
				Select the action $a_t$ based on the current policy $\pi_\vartheta(s_t)$. 
				
				Execute the action $a_t$, receive the reward $r_t$, and observe the updated state $s_{t+1}$ of the environment. 
				
				Store the tuple $(s_t, a_t, r_t, s_{t+1})$ in the experience replay buffer $\mathcal{M}$. 
			}
			
			\For{Training iteration k=1 to $K$}{
				Randomly sample several tuples $(s, a, r, s^{'})$ from the experience replay buffer $\mathcal{M}$. 
				
				For each tuple, compute the target Q-value $y$ using Equation (\ref{eq.Qv}).
				
				Update Q-networks parameters $\theta_1$, $\theta_2$ by minimizing equation (\ref{eq:Qpra}).
				
				Update the policy network parameters $\vartheta$ by minimizing equation (\ref{eq:Ppra}).
				
				Update the entropy coefficient $\lambda_e$ by minimizing Equation (\ref{eq:entc}). 
				
				Update the target Q-networks parameters $\theta^{'}_1$, $\theta^{'}_2$ by using the following equation:
				$\theta^{'}_i=\tau\theta_i+(1-\tau)\theta^{'}_i~,0<\tau<1~,i=1,2$.
			}
		}
	\end{algorithm}
	
	\subsection{Computational Complexity Analysis}
	
	For \textbf{Algorithm \ref{alg:HDBSCAN}}, the average time complexity of a single HDBSCAN clustering process is approximately $\mathcal{O}(K\log K)$, where $K$ denotes the total number of users. The complexity of assigning user clusters to UAVs is $\mathcal{O}(N_1^3)$, where $N_1=\max(N,U)$, and $U$ denotes the number of identified clusters. The time complexity of computing the attractive and repulsive forces generated by the UAVs based on their assigned cluster centers is $\mathcal{O}(N^2)$. In addition, this part also involves the computation of cluster centroids with a complexity of $\mathcal{O}(K)$, and the updates of the channel state and transmission rate, which require $\mathcal{O}(KM  N)$ computational cost. \par
	For \textbf{Algorithm \ref{alg:SAC}}, the computational time required by the SAC algorithm is primarily determined by the training of its policy network and value networks, both of which are Deep Neural Networks \cite{35}. The total time complexity for updating the parameters of the policy and value networks at each iteration can be expressed as 
	\begin{equation}
		\mathcal{O}_A = \mathcal{O}( N_B ( \sum_{i=1}^{I-1} l_i l_{i+1} + \sum_{j=1}^{J-1} \hat{l}_j l_{j+1} ) ),
	\end{equation}
	where $N_B$ denotes the batch size, $I$ and $J$ represent the number of layers in the value networks and policy network, respectively, and $l_i$ and $l_j$ denote the dimensions of the $i$ layer of the value networks and the $j$ layer of the policy network, respectively. The time complexity of the SAC algorithm during the training phase is expressed as $\mathcal{O}_t=\mathcal{O}(N_ETN_B\mathcal{O}_A)$,
	where $N_E$ denotes the number of training epoch, and $T$ represents the number of steps per epoch. During the execution phase, the agent makes decisions solely based on the trained policy. The time complexity of each decision is $\mathcal{O}_e=\mathcal{O}(\sum_{j=0}^{J-1} \hat{l}_j l_{j+1})$.
	
	\section{NUMERICAL RESULTS}
	\label{section:4}
	
	\subsection{Simulation Scenario and Parameter Setting}
	\begin{table}[htbp]
		\centering
		\caption{LIST OF SIMULATION PARAMETERS}
		\label{tab:1}
		\begin{tabular}{l|l}
			\toprule
			\textbf{Parameters} & \textbf{Value}\\
			\midrule
			Service area size &  500 m\\
			UAV maximum velocity ($V$) & 4 m/s\\
			UAV number ($N$) & 3\\
			UAV operating altitude ($z_n$) & 80--120 m\\
			UAV maximum transmitting power ($\mathrm{P}_{\mathrm{max}}$) & 30 dBm\\
			User maximum velocity & 0.8 m/s\\
			User number ($K$)& 12\\
			Communication user number ($K_c$) & 9\\
			Sensing target number ($K_s$) & 3\\
			Antenna number ($M$) & 4\\
			MA array maximum displacement ($D_{\mathrm{off}}$) & 0.625 m\\
			Antenna minimum separation distance ($D_{\mathrm{min}}$) & 0.0625 m\\
			Bandwidth ($\mathrm{B}$)& 20 MHz\\
			Carrier frequency & 2.4 GHz\\
			AWGN power ($\sigma^2$) & -110 dBm/Hz\\
			
			\hline
		\end{tabular}

	\end{table}
	
	In this section, we provide the performance evaluation of the proposed framework. In the scenario, we simulate the free movement of 9 communication users and 3 sensing targets within a square ground area of 500 meters on each side. Meanwhile, 3 UAVs are deployed in the air between 80 meters and 120 meters above the ground, with each UAV equipped with an MA array consisting of 4 antennas. Since the energy consumption of the main decision-making process is relatively low during the simulation, we did not consider this energy consumption in the current simulation. Other simulation parameters are shown in Table \ref{tab:1}.
	
	Additionally, we set the $\textrm{min\underline{~}cluster\underline{~}size}$ and $\textrm{min\underline{~}samples}$ to 2. Based on the number of UAVs and user groups, if these parameters are set too large, it may lead to too many users being classified as noise during the training process, thereby diminishing the clustering effect. The $\epsilon$ parameter is set to about $\textrm{10\%}$ of the scene's side length and represents the threshold for controlling the final cluster extraction in the algorithm. Based on the clusters generated in this step, the second step of the algorithm is executed. The agent generates the UAV's actual flight speed and direction based on the cluster assignment from the first step. Simultaneously, it generates the MA array adjustment strategy to assist in beam alignment, considering the spatial distribution characteristics of the associated objects.
	
	In the experiments, the proposed framework is compared with four alternative schemes. \textbf{Scheme 1:} The MA array configuration is retained as in the proposed design, however, no clustering algorithm is employed for air–ground association. This scheme is included to demonstrate the performance gain brought by the clustering mechanism within the overall framework. \textbf{Scheme 2:} All UAVs are equipped with FA arrays with uniform inter-element spacing, and no clustering algorithm is used for air–ground association. This scheme is introduced to highlight the performance improvement achieved by the MA array design in the considered scenario. \textbf{Scheme 3:} The MA array configuration is adopted, but the UAVs execute sensing and communication tasks along predefined fixed trajectories. This scheme is included to independently quantify the performance improvement contributed by the MA array. \textbf{Scheme 4:} All UAVs are equipped with FA arrays and follow fixed trajectories during task execution. This scheme represents a conventional UAV-assisted ISAC setup and serves as a lower performance bound to benchmark the gains of the other schemes.
		
	Additionally, we selected two representative DRL algorithms and an optimization algorithm based on evolutionary strategies as baseline comparisons in continuous control. The deep deterministic policy gradient (DDPG) algorithm, an off-policy method, is used to assess the advantages of SAC's maximum entropy mechanism in terms of exploration and avoiding local optima. The proximal policy optimization (PPO), an on-policy algorithm, is used to compare SAC's sample efficiency and training stability. Furthermore, to validate the effectiveness of the algorithm, we also compared the performance of the covariance matrix adaptation evolution strategy (CMA-ES) algorithm, which does not rely on gradient information and demonstrates strong performance in addressing high-dimensional, nonlinear, and non-convex optimization problems in continuous domain. To ensure the reliability of the comparison, the learning rate for all DRL algorithms is set to $10^{-4}$, with a hidden layer structure of [256, 256], a batch size of 256, and 1200 training epochs for each algorithm.
		
	\begin{figure}[t]
		\centering
		\includegraphics[width=\linewidth]{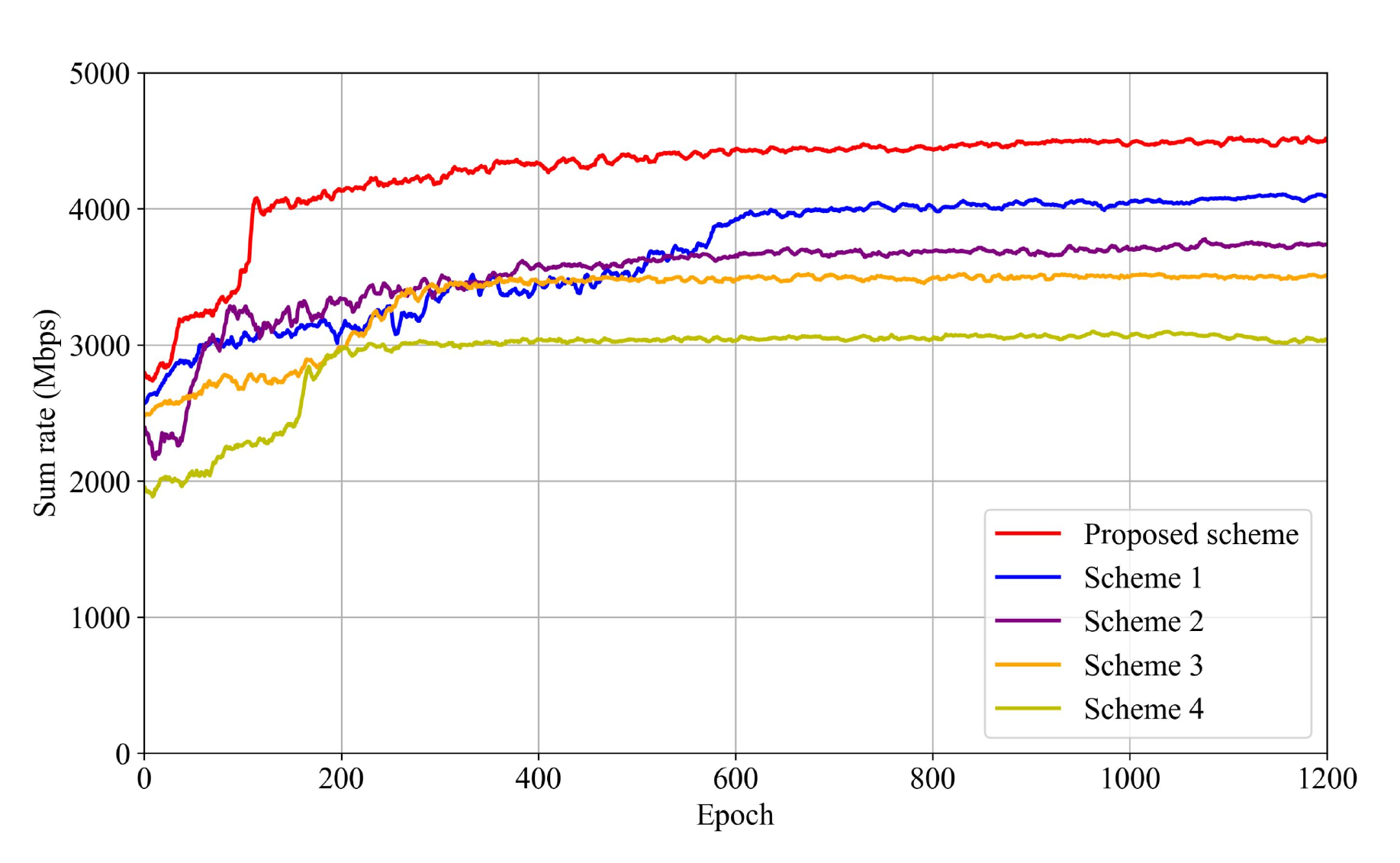}
		
		\caption{Sum rate under different schemes.}
		\label{t2}
	\end{figure}
	\begin{figure}[t]
		\centering
		\includegraphics[width=\linewidth]{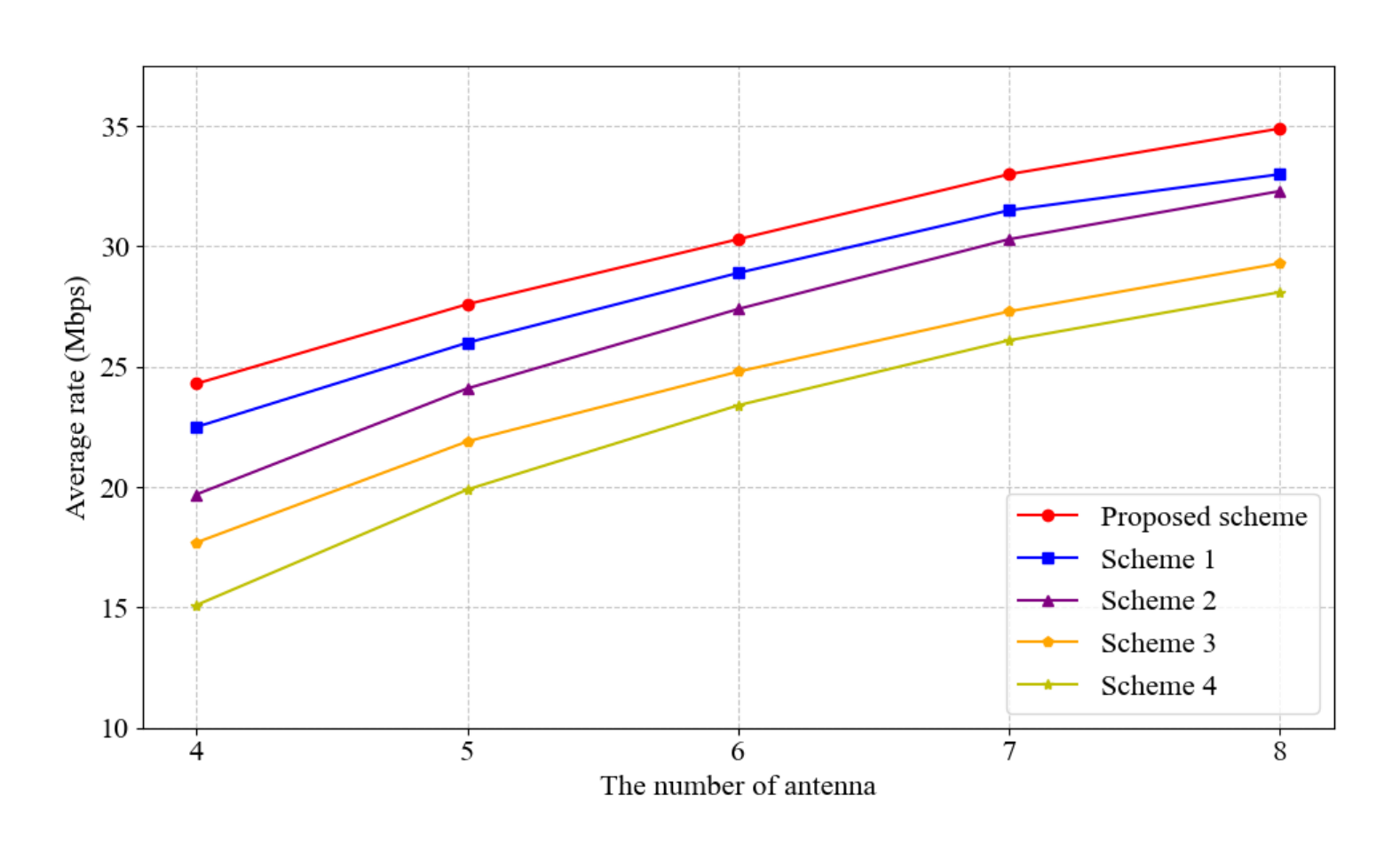}
		
		\caption{Average rate with respect to the number of antennas under different frameworks.}
		\label{t3}
	\end{figure}
	\subsection{Performance of Proposed Algorithms}
	\begin{figure}[t]
		\centering
			\centering
			\includegraphics[width=0.9\linewidth]{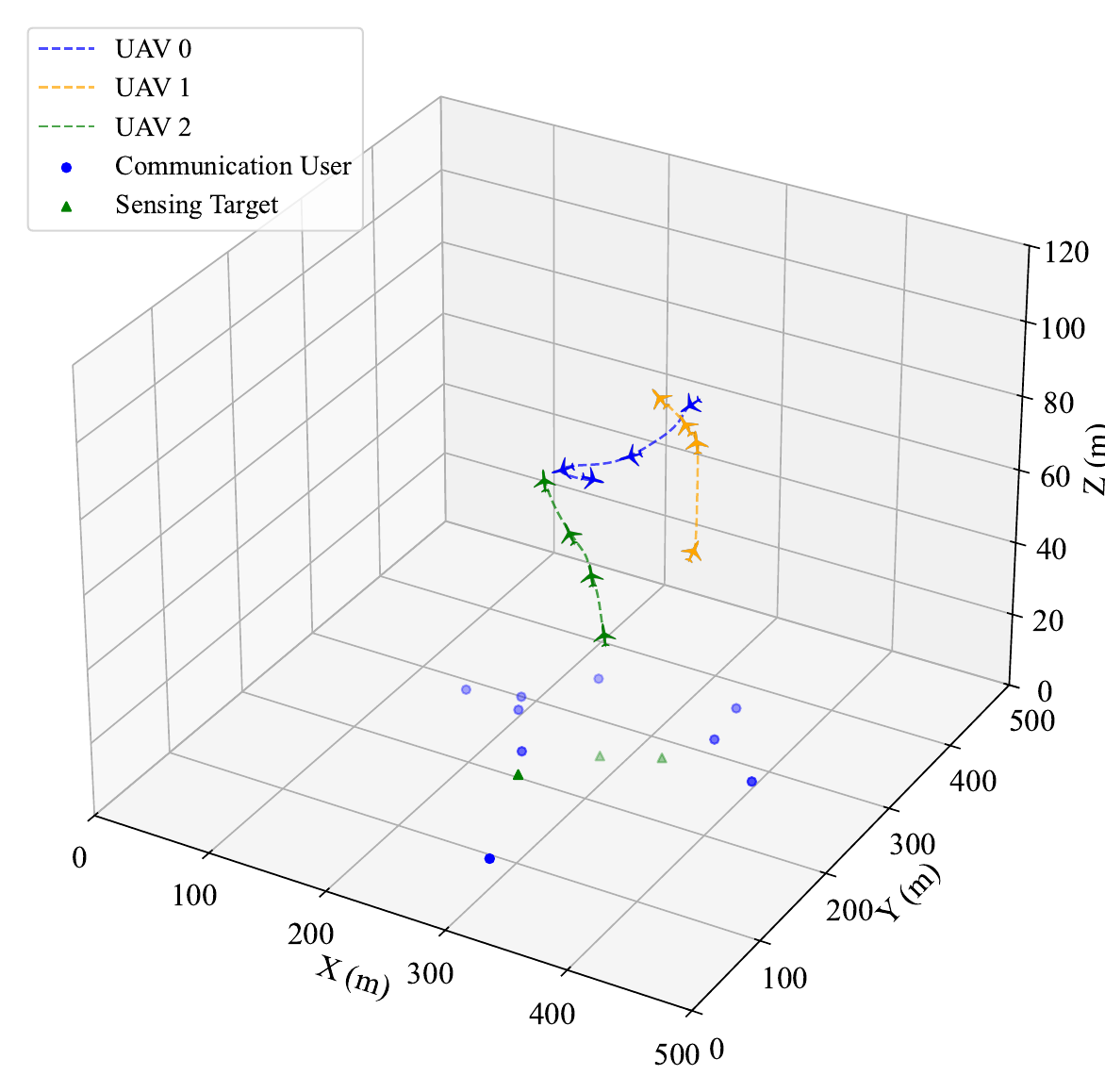}
			
			\caption{Trajectories of UAVs based on our algorithm.}
			\label{t4a}
	\end{figure}
	
	Fig. \ref{t2} illustrates the performance comparison between the proposed scheme and four alternative design schemes. The results show that the proposed scheme outperforms the other schemes in both overall performance and early-stage exploration efficiency. When comparing the proposed scheme with \textbf{Scheme 1} and \textbf{Scheme 2}, it is evident that embedding the MA array in this scenario allows for a higher performance ceiling in the later training stages. However, due to the increased dimensionality from the array movements, \textbf{Scheme 1} shows slower exploration in the early stages compared to the other two schemes. Furthermore, observing the three curves reveals that the integration of the clustering algorithm in this optimization framework improves overall performance. The clustering behavior reduces the exploration pressure on the agent, which is evident in the faster early-stage exploration rate. Additionally, a comparison between \textbf{Scheme 3} and \textbf{Scheme 4} reveals that even with the slower exploration, the MA array attains a higher performance ceiling than the traditional FA array in the absence of any optimization. Overall, every component in the proposed optimization framework is closely integrated. The combination of the MA array and the clustering algorithm leads to a significant performance improvement.

	Fig. \ref{t3} illustrates the average communication performance per time slot for the different schemes as the number of antennas increases. It can be observed that the proposed solution achieves the best overall performance, with the clustering algorithm offering a slight improvement in communication efficiency. The performance curves show that the use of an MA array significantly enhances system performance, though this advantage diminishes as the number of antennas grows. This can be attributed to the UAV's load constraints: as the number of antennas increases, the UAV's mobility range is limited, making the performance gain less significant compared to that of the fixed antenna array. Additionally, increasing the number of antennas adds more action dimensions to the agent's strategy, thereby resulting in higher computational overhead and diminishing the overall benefit.
	
	Fig. \ref{t4a} shows the 3D trajectory of the UAV, where ground users are randomly distributed on the surface. The UAV’s flight path is jointly determined by the clustering algorithm and the policy, with the trajectory continuously adjusting in response to the users' random movements. Both the UAV's flight direction and speed change over time. Additionally, the flight strategy account for factors such as cluster hotspot tracking and beamforming, which results in a more complex and winding trajectory.

	\begin{figure}[t]
		\centering
		\includegraphics[width=\linewidth]{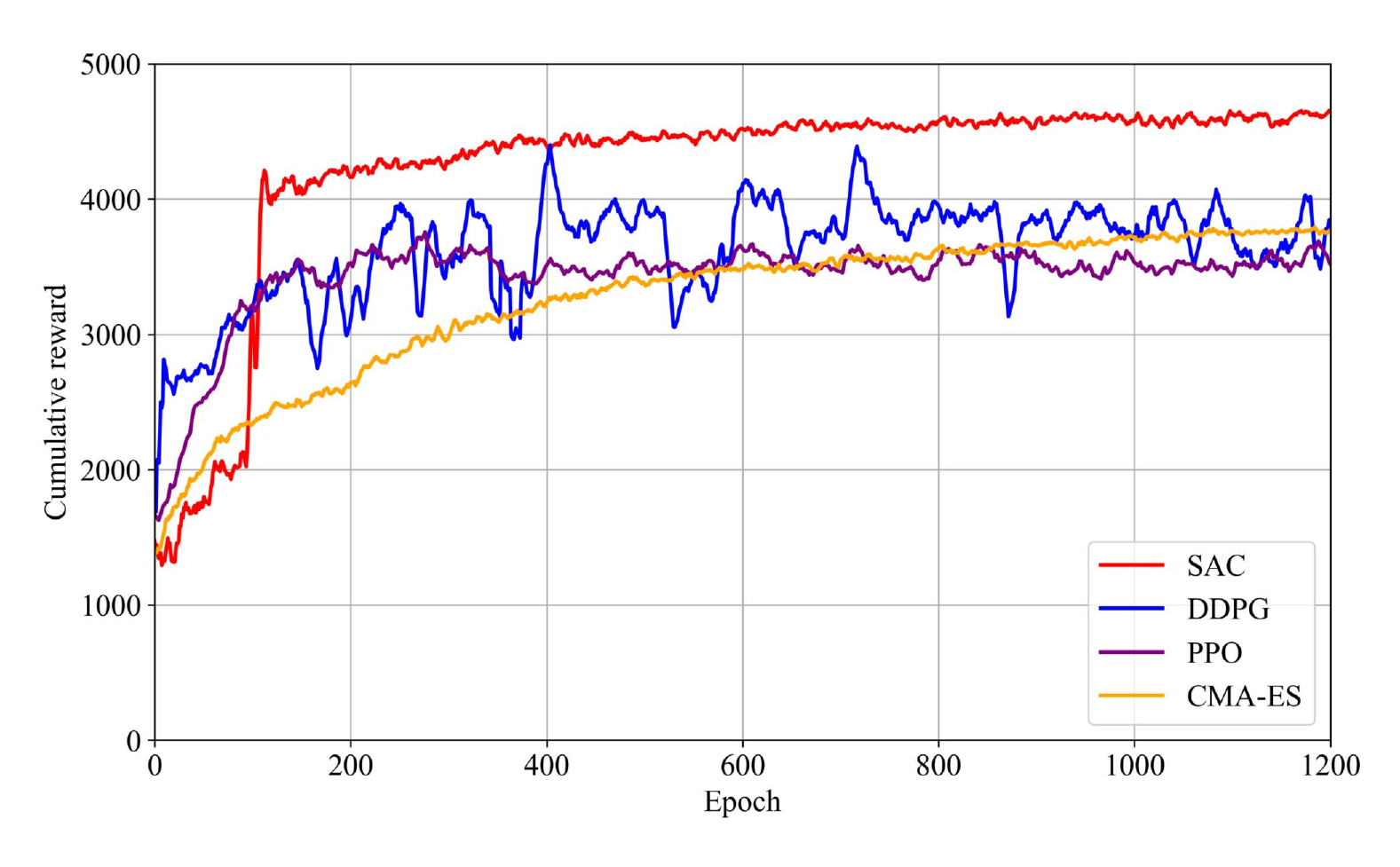}

		\caption{Cumulative reward under different algorithms.}
		\label{t5}
	\end{figure}

	\begin{figure}[t]
		\centering
		\includegraphics[width=\linewidth]{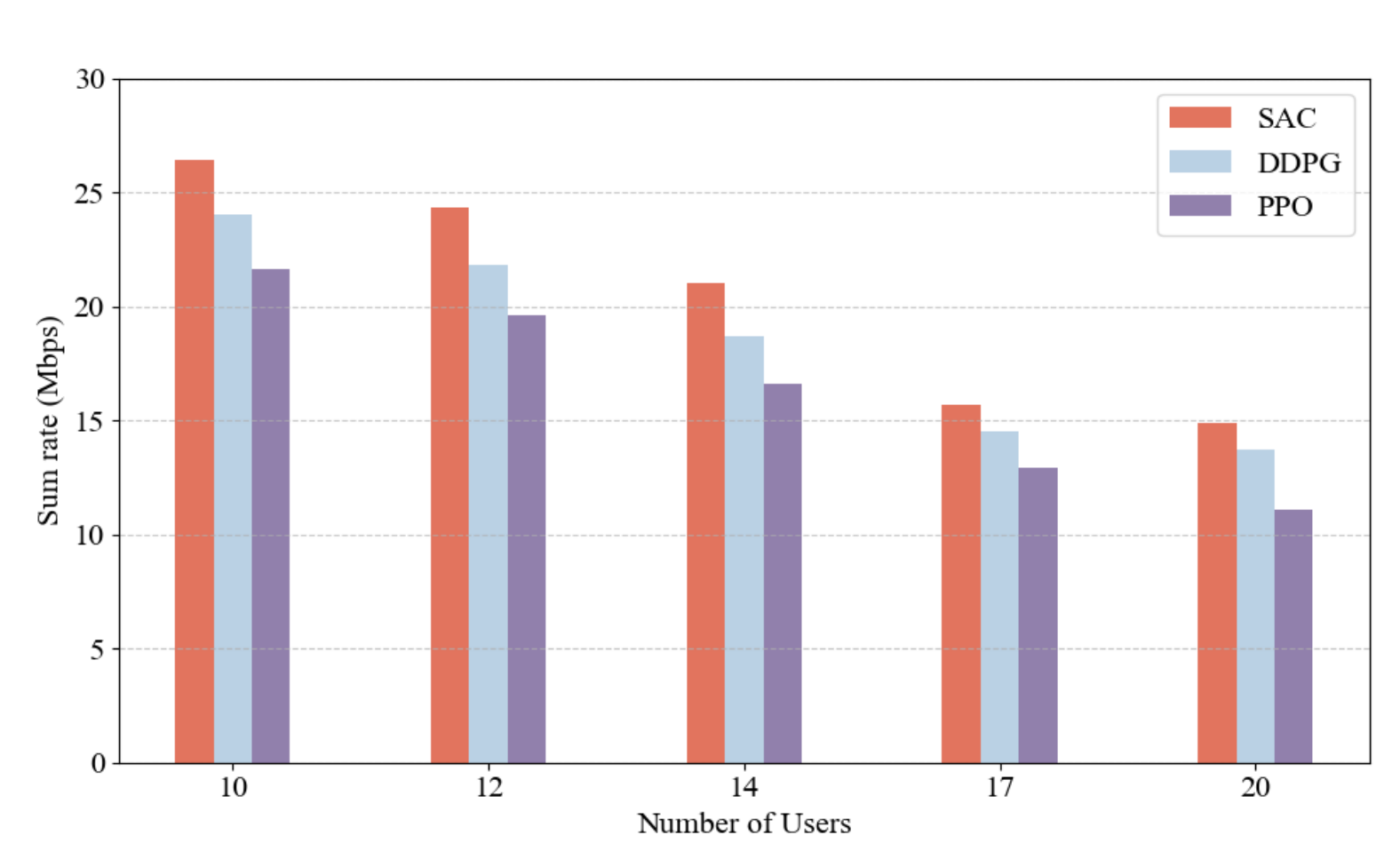}
		\caption{Sum rate under different number of users.}
		\label{t7}
	\end{figure}
	\subsection{Performance Comparison of Different Algorithms}
	
	Fig. \ref{t5} compares the cumulative reward under different algorithms. The cumulative reward is the sum of all rewards accumulated over the training steps within one epoch. As shown, the cumulative rewards for the four algorithms, from highest to lowest, are SAC, DDPG, CMA-ES and PPO. The cumulative reward achieved by SAC is approximately 800 higher than that of DDPG at the final stage. In terms of the training process, the SAC algorithm achieves an effective training strategy after approximately 130 epochs of exploration, allowing for a rapid performance improvement. It then enters an entropy annealing phase, where the strategy stabilizes and gradually increases. The total training time is 4 hours and 51 minutes. The DDPG algorithm experiences significant data fluctuations during training, with the strategy being less stable than SAC. However, the amplitude of these fluctuations diminishes over time, and the total training duration is 5 hours and 5 minutes. The CMA-ES algorithm maintains a steady upward trend, but it takes approximately 600 steps to reach the plateau phase, making it less efficient compared to the other algorithms. The PPO algorithm maintains a stable training strategy, with the performance entering a plateau after about 200 epochs, and the growth in reward becomes more gradual. The total training time for PPO is 6 hours and 12 minutes. These results indicate that the SAC algorithm, when handling the high-dimensional continuous action space in this scenario, better balances exploration and exploitation, significantly outperforming the other three algorithms.
	
	\begin{figure}[t]
		\centering
		\includegraphics[width=\linewidth]{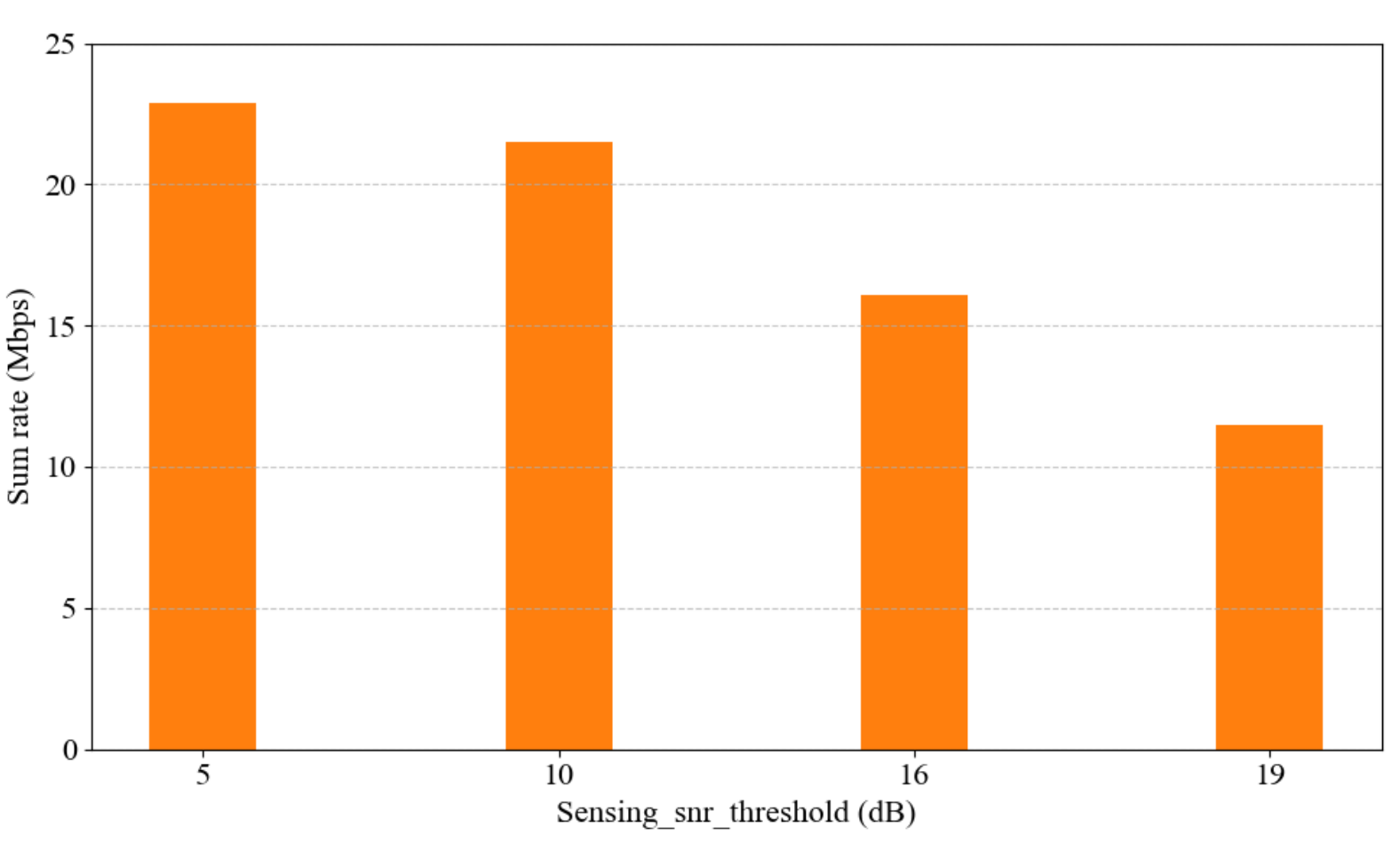}
		\caption{Sum rate with respect to changes in the sensing threshold.}
		\label{t8}
	\end{figure}

	Fig. \ref{t7} illustrates the impact of changes in the number of users on algorithm performance. The figure presents the average communication rate per time slot, calculated after the algorithms have converged. The number of UAVs is fixed at 3, and the number of sensing targets is also set to 3. In this case, the variation in the number of users is primarily due to changes in the number of communication users. As shown in the figure, the communication performance of the ISAC system proposed in this paper decreases as the number of communication users increases. This is attributed to the limited resources available on the UAVs. As the number of communication users grows, the number of users that each UAV must serve also increases, which leads to a reduction in the resources allocated to each user, resulting in a lower SINR. However, the proposed solution maintains a certain level of stability even in dense user populations, as the increase in the number of users does not cause a sharp decline in performance.
	
	Fig. \ref{t8} shows the communication performance under different sensing SINR thresholds. Since sensing tasks in the ISAC system consume a portion of the UAV’s resources, thereby reducing the available communication resources, the communication performance of the ISAC system decreases as the demand for sensing tasks increases.

	\section{Conclusion}
	\label{section:5}
	This paper has investigated an ISAC scenario in which multiple UAVs equipped with MA arrays are deployed in 3D space to serve ground users with random mobility. Our work has integrated MA arrays into multi-UAV ISAC systems and has introduced a user re-association scheme based on the HDBSCAN clustering algorithm to ensure rational air-to-ground associations. Furthermore, an SAC-based deep reinforcement learning approach has been proposed to jointly optimize UAV trajectories, beamforming vectors, and antenna placements, aiming to maximize the total data rate while maintaining the required sensing signal-to-noise ratio. Extensive simulation results have demonstrated the superiority of MA arrays over fixed-position antenna arrays in the considered scenario, validated the performance enhancement achieved by the HDBSCAN clustering algorithm, and confirmed the stability and effectiveness of the proposed method under varying system configurations.
	

\end{document}